\documentclass[aps,print,a4paper,12pt,showpacs]{revtex4}%
\usepackage{amsfonts}
\usepackage{amsmath}
\usepackage{amssymb}
\usepackage{graphicx}%
\begin{document}
\title{Soliton solution of continuum magnetization-equation in conducting ferromagnet
with a spin-polarized current}
\author{Z. D. Li$^{1}$, J. Q. Liang$^{1}$, Lu Li$^{1}$, and W.-M Liu$^{2}$}
\affiliation{$^{1}${Institute of Theoretical Physics and Department of Physics, Shanxi
University, Taiyuan 030006, China}}
\affiliation{$^{2}$National Lab. of Magnetism, Institute of Physics, Chinese Academy of
Sciences, Beijing 100080, China}

\begin{abstract}
Exact soliton solutions of a modified Landau-Lifshitz equation for the
magnetization of conducting ferromagnet in the presence of a spin-polarized
current are obtained by means of inverse scattering transformation. From the
analytical solution effects of spin-current on the frequency, wave number, and
dispersion law of spin wave are investigated. The one-soliton solution
indicates obviously current-driven precession and periodic shape-variation as
well. The inelastic collision of solitons by which we mean the shape change
before and after collision appears due to the spin current. We, moreover, show
that complete inelastic collisions can be achieved by adjusting spectrum and
current parameters. This may lead to a potential technique for shape control
of spin wave.

\end{abstract}

\maketitle

Considerable attention has been paid to the dynamics of magnetization
associated with spin-polarized current in layered materials and in Mn oxides
recently. Both theoretical and experimental investigations mainly concentrated
on large magnetoresistance are of fundamental importance in the understanding
of magnetism and applied interest in the fabrication of magnetic devices.
Spin-transfer from spin-polarized current to magnetization of conducting
ferromagnetic films is one of intriguing features which is theoretically
proposed in Ref. \cite{Slonczewski,Berger} and subsequently
verified\cite{Katine,Bazaliy} in experiment. Many
studies\cite{Sun,Waintal,Stile,Zhang,Murakami} on this phenomenon have been
followed since the spin-transfer mechanism was first explained conceptually.
However, the dynamics of magnetization in the presence of spin current has not
been well understood. In Ref. \cite{Bazaliy} a continuum equation for the
magnetization of conducting ferromagnet in the presence of spin-polarized
current is derived and is seen to be a modified Landau-Lifshitz equation with
an additional topological term. The spatial dependence of the magnetization is
evaluated \cite{Bazaliy} and several solutions of one-dimension are also
discussed. We in this paper study the soliton solution of the modified
Landau-Lifshitz equation given in Ref. \cite{Bazaliy} from which the current
induced precession of the magnetization and soliton-soliton collisions are
investigated. The effect of spin-current on the dynamics of magnetization is
demonstrated explicitly. We obtain exact soliton solutions by means of inverse
scattering transformation in one-dimensional geometry. An intriguing feature
is that inelastic collisions generally appear due to the spin-polarized
current and the complete inelastic collisions which may lead to a interesting
technique of soliton filter and switch can be achieved in special cases by
adjusting of the parameters of spin-polarized current.

Following Ref. \cite{Bazaliy} we consider a current propagating through
conducting ferromagnet and assume that the conducting electrons interact only
with the local magnetization $\mathbf{M}$. A continuum equation for the
magnetization is obtained in the local magnetization frame as a
Landau-Lifschitz type\cite{Bazaliy},
\begin{equation}
\frac\partial{\partial t}\mathbf{M}=\mathbf{M}\times\widetilde{J}%
\mathbf{M}_{zz}-\gamma\mathbf{M}_{z}, \label{LL}%
\end{equation}
where $\mathbf{M}(z,t)=(M^{x}(z,t)$, $M^{y}(z,t)$, $M^{z}(z,t))$ is the local
magnetization and $\widetilde{J}=\frac{g\mu_{B}M}\hbar J$ with $J$ being the
exchange interacting constant between the local magnets. The parameter
$\gamma$ describes effect of the current which is the same as defined in Ref.
\cite{Bazaliy}. The length of magnetization vector is set to unit ,
$\mathbf{M}^{2}\left(  z,t\right)  =1$, for the sake of simplicity. We begin
with the inverse scattering transformation which is a useful method to solve
the nonlinear equation (\ref{LL}). By means of Ablowitz-Kaup-Newell-Segur
method one can construct Lax equations for the Eq. (\ref{LL}) as
\begin{align}
\frac\partial{\partial z}\Psi\left(  z,t,\lambda\right)   &  =U_{1}\left(
\lambda\right)  \Psi\left(  z,t,\lambda\right)  ,\nonumber\\
\frac\partial{\partial t}\Psi\left(  z,t,\lambda\right)   &  =U_{2}\left(
\lambda\right)  \Psi(z,t,\lambda), \label{lax1}%
\end{align}
where $\lambda$ is a spectrum parameter, $\Psi\left(  z,t,\lambda\right)  $ is
eigenfunction corresponding to the spectrum $\lambda$. The operators
$U_{1}\left(  \lambda\right)  $ and $U_{2}\left(  \lambda\right)  $ are given
in the following form
\begin{align}
U_{1}  &  =-i\widetilde{J}\left(  \lambda-\lambda_{0}\right)  \left(
M\cdot\sigma\right)  ,\nonumber\\
U_{2}  &  =i2\widetilde{J}^{3}\left(  \lambda^{2}-\lambda_{0}^{2}\right)
\left(  M\cdot\sigma\right) \nonumber\\
&  \mathbf{-}\widetilde{J}^{2}\left(  \lambda-\lambda_{0}\right)  \left(
M\cdot\sigma\right)  \left(  M_{z}\cdot\sigma\right)  , \label{lax2}%
\end{align}
where the real parameter $\lambda_{0}=\gamma(4\widetilde{J}^{2})^{-1}$
indicates the effect of the current and $\mathbf{\sigma}$ is Pauli matrix.
Thus Eq. (\ref{LL}) can be recovered from the compatibility condition
$\frac\partial{\partial t}U_{1}-\frac\partial{\partial z}U_{2}+\left[
U_{1},U_{2}\right]  =0$. We consider the following natural boundary condition
at initial time($t=0$), $\mathbf{M}\left(  z\right)  \equiv\left(  M^{x}%
,M^{y},M^{z}\right)  \rightarrow\left(  0,0,1\right)  \text{ as}\left|
z\right|  \rightarrow\infty$. We then have the asymptotic form of Eq.
(\ref{lax1}) at $\left|  z\right|  \rightarrow\infty$,
\begin{equation}
\partial_{z}E(z,\lambda)=L_{0}(\lambda)E(z,\lambda),
\end{equation}
where
\begin{equation}
E(z,\lambda)=e^{-i\widetilde{J}\left(  \lambda-\lambda_{0}\right)  z\sigma
_{3}},\ \ \ L_{0}(\lambda)=-i\widetilde{J}\left(  \lambda-\lambda_{0}\right)
\sigma_{3},
\end{equation}
Based on the Lax equations (\ref{lax1}), we can derive the exact solution of
$N$-soliton trains by employing of the inverse scattering
transformation\cite{zdli,Ablowitz}. As a special case that $N=1$ the exact
one-soliton solution is given as follows:
\begin{align}
M^{x}  &  =\frac1{\Delta_{1}}\left[  -2\left(  \alpha_{1}-\lambda_{0}\right)
\beta_{1}\sin\left(  \Phi_{1}-\phi_{1}\right)  \cosh\Theta_{1}\right.
\nonumber\\
&  -2\beta_{1}^{2}\cos\left(  \Phi_{1}-\phi_{1}\right)  \sinh\Theta_{1}\left.
{} \right]  ,\nonumber\\
M^{y}  &  =\frac1{\Delta_{1}}\left[  2\left(  \alpha_{1}-\lambda_{0}\right)
\beta_{1}\cos\left(  \Phi_{1}-\phi_{1}\right)  \cosh\Theta_{1}\right.
\nonumber\\
&  \left.  -2\beta_{1}^{2}\sin\left(  \Phi_{1}-\phi_{1}\right)  \sinh
\Theta_{1} \right]  ,\nonumber\\
M^{z}  &  =1-\frac{2\beta_{1}^{2}}{\Delta_{1}}, \label{onesoliton1}%
\end{align}
where
\begin{align*}
\Delta_{1}  &  =|\lambda_{1}-\lambda_{0}|^{2}\cosh^{2}\Theta_{1},\\
\Theta_{1}  &  =2\widetilde{J}\beta_{1}(z-V_{1,M}t)-z_{1},\\
\Phi_{1}  &  =2\widetilde{J}\left(  \alpha_{1}-\lambda_{0}\right)
z-4\widetilde{J}^{3}\left[  \alpha_{1}^{2}-\beta_{1}^{2}-\lambda_{0}%
^{2}\right]  t-\phi_{1}.
\end{align*}
Parameter $V_{1,M}=4\widetilde{J}^{2}\alpha_{1}$ denotes the velocity of
envelope, $z_{1}=\ln[(2\widetilde{J}\beta_{1})^{-1}c_{1}]$ is the center
position, and $\phi_{1}=\arg\left[  \widetilde{J}\left(  \lambda_{1}%
-\lambda_{0}\right)  \right]  $ is the initial phase of the spin wave. The
parameter $\lambda_{1}=\alpha_{1}+i\beta_{1}$ denotes eigenvalue with
$\alpha_{1}$, $\beta_{1}$ being the real and imaginary parts respectively, and
$c_{1}$ is a real constant of integration. The solution Eq. (\ref{onesoliton1}%
) describes a current-driven precession of magnetization with periodic shape
variation. The center of solitary wave moves with velocity $V_{1,M}$, while
the wave amplitude and width vary periodically with time. We see that the
spin-polarized current imparts a torque to the magnetization due to local
exchange interaction between electron-spin and the magnetic moment. This
observation is in accord with the prediction in
Ref.\cite{Slonczewski,Berger,Myers,Bazaliy,Zhang}. As a consequence of
reaction the current flow is strongly affected by the orientation of the
magnetic moments. Thus a higher electrical resistance in magnetic layer may
occur. The spin-polarized current can be used to adjust the precession of
magnetic moment and the wave shape as well. We then provide in principle a
mechanism of current-control of the spin wave.

To see closely the physical significance of one-soliton solution it is helpful
to show the parameter-dependence of Euler angles of the magnetization vector
which in a spherical coordinate is written as $\mathbf{M}\left(  z,t\right)
\equiv\left(  \sin\theta\cos\varphi,\sin\theta\sin\varphi,\cos\theta\right)
$. From the Eq. (\ref{onesoliton1}) we find
\begin{equation}
\cos\theta=1-\frac{A_{M}}{\cosh^{2}\left[  \digamma_{1}^{-1}(z-V_{1,M}%
t)-z_{1}\right]  }, \label{polar1}%
\end{equation}
\begin{equation}
\varphi=\frac\pi2-\phi_{1}+k_{1}z-\Omega_{1}t+\arctan\left(  \frac{\beta_{1}%
}{\alpha_{1}-\lambda_{0}}\tanh\Theta_{1}\right)  , \label{polar3}%
\end{equation}
where $A_{M}=\frac{2\beta_{1}^{2}}{|\lambda_{1}-\lambda_{0}|^{2}}$,
$\digamma_{1}=\frac1{2\widetilde{J}\beta_{1}}$ are amplitude and width of the
soliton respectively. The wave number is $k_{1}=k_{0}-k_{S}$ with
$k_{0}=2\widetilde{J}\alpha_{1}$ denoting the wave number in the absence of
the current while $k_{S}=2\widetilde{J}\lambda_{0}$ is the wave number shift
induced by the spin-polarized current. The frequency of magnetization
precession is seen to be $\Omega_{1}=\Omega_{0}-\Omega_{S}$ with $\Omega
_{0}=4\widetilde{J}^{3}\left(  \alpha_{1}^{2}-\beta_{1}^{2}\right)  $ being
the frequency in the absence of current and $\Omega_{S}=4\widetilde{J}%
^{3}\lambda_{0}^{2}$ the frequency shift induced by spin-polarized current. We
see that the effect of current reduces both the wave number and frequency. For
a large enough current such that $\lambda_{0}^{2}>\alpha_{1}^{2}-\beta_{1}%
^{2}$ an instability occurs \cite{Bazaliy}. We can rewrite the frequency, i.e.
the energy spectrum as
\begin{equation}
\Omega_{1}=\widetilde{J}k_{1}\left(  k_{1}+2k_{S}\right)  -4\widetilde{J}%
^{3}\beta_{1}^{2}. \label{polar4}%
\end{equation}
We then see that in the absence of current the minimum of the energy spectrum
i.e. $\Omega_{0,\min}=0$ is located at $k_{0,\min}=\sqrt{4\widetilde{J}%
^{2}\beta_{1}^{2}}$ while the current shifts the position of minimum by an
amount $\delta=\sqrt{k_{S}^{2}+4\widetilde{J}^{2}\beta_{1}^{2}}-(k_{S}%
+\sqrt{4\widetilde{J}^{2}\beta_{1}^{2}})$. In the absence of spin-polarized
current, i.e. $\lambda_{0}=0$, the solution Eqs. (\ref{polar1}) (\ref{polar3})
reduces to the soliton solution in an isotropic spin chain \cite{Takhtajan}.

In the limit case that the amplitude $A_{M}$ approaches zero, namely
$\beta_{1}\rightarrow0$, the soliton width $\digamma_{1}$ diverges, the
envelope velocity $V_{1,M}$ attains its maximum value $2\sqrt{\widetilde
{J}\Omega_{0}}$, and the solution shown in Eq. (\ref{polar1}) and Eq.
(\ref{polar3}) takes the form such that
\[
M^{z}\rightarrow1,\text{ }\varphi\rightarrow\frac\pi2-\phi_{1}+k_{1}%
z-\Omega_{1}t
\]
indicating a small linear solution of magnon. In this case the quadratic
dispersion law is seen to be $\Omega_{1}=\widetilde{J}\left(  k_{0}^{2}%
-k_{S}^{2}\right)  =\widetilde{J}k_{1}\left(  k_{1}+2k_{S}\right)  $. We also
notice that the phase velocity of the precession is $\frac{\Omega_{1}}{k_{1}%
}=\frac{V_{1,M}}2+\frac{V_{S}}2$ which possesses a correction value
$\frac{V_{S}}2$ which is the half of envelope velocity. Whereas the group
velocity of precession $\frac{d\Omega_{1}}{dk_{1}}=V_{1,M}$ coincides with the
envelope velocity.

We now consider another special case of the general $N$-soliton trains, i.e.,
the two-soliton solution which is seen to be
\begin{align}
M^{x}  &  =\operatorname{Re}\left[  -i2\Gamma_{2}(1-i\Gamma_{1})\right]
,\nonumber\\
M^{y}  &  =\operatorname{Im}\left[  i2\Gamma_{2}(1-i\Gamma_{1})\right]
,\nonumber\\
M^{z}  &  =\left\vert 1-i\Gamma_{1}\right\vert ^{2}-\left\vert \Gamma
_{2}\right\vert ^{2}, \label{twosoliton1}%
\end{align}
where
\begin{align*}
\Gamma_{1}  &  =\frac{1}{W}\left[  \left(  g_{1}-g_{3}\right)  g_{6}+\left(
g_{2}-g_{4}\right)  g_{5}\right]  ,\\
\Gamma_{2}  &  =\frac{1}{W}\left[  -(\overline{g_{1}}-\overline{g_{3}}%
)g_{8}-(\overline{g_{2}}-\overline{g_{4}})g_{7}\right]  ,
\end{align*}
with
\begin{align*}
g_{1}  &  =1+\left\vert q_{1}\right\vert ^{2}+\chi_{1}\overline{\chi}_{2}%
q_{1}\overline{q}_{2},\\
g_{2}  &  =1+\left\vert q_{2}\right\vert ^{2}+\overline{\chi}_{1}\chi
_{2}\overline{q}_{1}q_{2},\\
g_{3}  &  =\overline{\chi}_{1}\left\vert q_{1}\right\vert ^{2}+\chi_{1}%
q_{1}\overline{q}_{2},\text{ }\\
g_{4}  &  =\overline{\chi}_{2}\left\vert q_{2}\right\vert ^{2}+\chi
_{2}\overline{q}_{1}q_{2},\\
g_{5}  &  =-\xi_{1}\left\vert q_{1}\right\vert ^{2}-\chi_{1}\xi_{2}%
q_{1}\overline{q}_{2},\\
g_{6}  &  =-\xi_{2}\left\vert q_{2}\right\vert ^{2}-\chi_{2}\xi_{1}%
\overline{q}_{1}q_{2},\\
g_{7}  &  =-\xi_{1}\overline{q}_{1},\text{ }\\
g_{8}  &  =-\xi_{2}\overline{q}_{2},
\end{align*}%
\begin{align*}
\chi_{1}  &  =\frac{2\beta_{1}\left(  \lambda_{1}-\lambda_{0}\right)
}{-i\left(  \lambda_{1}-\overline{\lambda}_{2}\right)  \left\vert \lambda
_{1}-\lambda_{0}\right\vert },\\
\chi_{2}  &  =\frac{2\beta_{2}\left(  \lambda_{2}-\lambda_{0}\right)
}{-i\left(  \lambda_{2}-\overline{\lambda}_{1}\right)  \left\vert \lambda
_{2}-\lambda_{0}\right\vert },\\
W  &  =g_{1}g_{2}-g_{3}g_{4},\\
q_{j}  &  =e^{-\Theta_{j}+i\Phi_{j}},\text{ }\\
\xi_{j}  &  =2\beta_{j}\left\vert \lambda_{j}-\lambda_{0}\right\vert ^{-1},
\end{align*}
and
\begin{align}
\Theta_{j}  &  =2\widetilde{J}\beta_{j}(z-V_{j,M}t)-z_{j},\nonumber\\
\Phi_{j}  &  =k_{j}z-\Omega_{j}t-\phi_{j}, \label{twosoliton2}%
\end{align}
where $V_{j,M}=4\widetilde{J}^{2}\alpha_{j}$ denotes the velocity of envelope
, $z_{j}=\ln[(2\widetilde{J}\beta_{j})^{-1}c_{j}]$ the center position,
$\phi_{j}=\arg\left[  \widetilde{J}\left(  \lambda_{j}-\lambda_{0}\right)
\right]  $ the initial phase, $k_{j}=2\widetilde{J}\left(  \alpha_{j}%
-\lambda_{0}\right)  $ the wave number, and $\Omega_{j}=4\widetilde{J}%
^{3}\left[  \alpha_{j}^{2}-\beta_{j}^{2}-\lambda_{0}^{2}\right]  $ is
frequency. The parameter $\lambda_{j}=\alpha_{j}+i\beta_{j}$ is eigenvalue
parameter, and $c_{j}$ is real constant of integration, $j=1,2$. The solutions
(\ref{twosoliton1}) describe in general a inelastic scattering process of two
solitary waves with different center velocities and different shape-variation
frequencies. Before collision, the two solitons move towards each other, one
with velocity $V_{1}$ and shape variation frequency $\Omega_{1}$, the other
with $V_{2}$ and $\Omega_{2}$ respectively. The interaction potential between
two solitons is a complicated function of current-dependent parameter
$\lambda_{0}$ and eigenvalue $\lambda_{j}$. For the case that $\alpha
_{j}=\beta_{j}$, the shape-variation frequencies $\Omega_{j}(j=1,2)$ of
two-soliton depend only on the parameters of spin-polarized current seen from
Eq.\ (\ref{twosoliton2}). In the case of $\lambda_{0}=0$, the solutions
(\ref{twosoliton1}) reduce to that of the usual two-soliton solution with two
center velocities while without shape change \cite{Takhtajan}. A interesting
process in the absence of spin-polarized current is that the collision can
result in the interchange of amplitude $A_{j}$ and phase $\Phi_{j}(j=1,2)$
like exactly in the case of elastic collision of two particles
\cite{Takhtajan}.

It is interesting to show the inelastic collision graphically. The head on
collision is explained in Fig. 1 and Fig. 2 for suppressed amplitudes of
$M_{1} $ and $M_{2}$ respectively after collision. This result shows that we
may adjust the incoming spin current and the spectral parameters to control
the shape of soliton of the magnetization. The dissipationless quantum spin
current at room temperature reported in Ref. \cite{Murakami} may be used to
realize experimentally the soliton-control in future. Our theoretical
observations predict the magnetic random-access memories in which the memory
elements are controlled by local exchange-effect forces induced by
spin-polaried current rather than by long-range magnetic fields.

In terms of inverse scattering transformation the exact soliton solutions for
the magnetization in conducting ferromagnet in the presence of a
spin-polarized current are obtained. Our solutions predict two intrinsic
features of the effect of spin-polarized current on the magnetization: (1)
Spin-polarized current induces the precession and shape variation of the
solitary waves of magnetization. (2) The inelastic collision of solitons. The
effect of spin-polarized current on the magnetization is similar to that of
the periodically time-varying external magnetic fields reported earlier
\cite{zdli} and is in agreement with the observations in Refs.
\cite{Slonczewski,Berger,Myers,Bazaliy,Zhang}.

This work was supported by National Natural Science Foundation of China under
Grant Nos. 10075032, 10174095, 90103024 and provincial overseas scholar
foundation of Shanxi.

Figure caption

Fig. 1 Inelastic head on collision expressed by Eq. (\ref{twosoliton1}) when
$M_{1}$ suppressed, where $\lambda_{1}=-0.4-i0.3$, $\lambda_{2}=0.5+i0.45$,
$\widetilde{J}=0.9$, $c_{1}=-0.55$, $c_{2}=3.8$, $\lambda_{0}=0.2$.

Fig. 2 Inelastic head on collision expressed by Eq. (\ref{twosoliton1}) when
$M_{2}$ suppressed, where $\lambda_{1}=-0.45+i0.3$, $\lambda_{2}=0.52-i0.47$,
$\widetilde{J}=0.9$, $c_{1}=0.55$, $c_{2}=-3.8$, $\lambda_{0}=0.2$.

\end{document}